\documentclass[journal] {IEEEtran}

\usepackage{bm}
\usepackage{array}
\usepackage[cmex10]{amsmath}
\usepackage{amsfonts}
\usepackage{verbatim}
\usepackage[dvipdfmx]{graphicx} 
\usepackage{epsfig}
\usepackage{amsthm}
\usepackage[caption=false,font=footnotesize]{subfig}
\usepackage{fixltx2e}
\usepackage{rotating}

\usepackage{url}
\usepackage{cite}
\newtheorem{thm}{Theorem}

\usepackage[normalem]{ulem}
\usepackage{hyperref}
\usepackage{color,array,colortbl}
\definecolor{mygray}{gray}{.9}

\ifCLASSINFOpdf
\else
\fi

\begin{document}
%
\title{Minimax Design of Nonlinear Phase FIR Filters with Optimality Certificates}
%
%
%

\author{Sefa~Demirtas \thanks{The author is with Analog Devices Lyric Labs. Email (work): sefa.demirtas@analog.com, (personal): sefa@alum.mit.edu.}
}

\maketitle

\begin{abstract}

The Parks-McClellan algorithm provides an efficient method for designing a linear phase FIR filter with a pre-specified weight function on the approximation error. For the given filter order and the specified weight function, the filter designed with this algorithm will have the unique optimal frequency response that approximates a desired filter response as certified by the alternation theorem. In this paper, a nonlinear phase FIR filter design algorithm is provided that allows the specification of a piecewise constant weight function on the approximation error in an analogous manner to linear phase FIR filters. For the given filter order and weight function, the resulting filter will provably have the unique optimal magnitude response that approximates a desired filter response, where the certification of optimality is given and is also based on the alternations that the weighted error function exhibits. Furthermore, the method is applicable to designing filters with both real- and complex-valued coefficients, which in turn determines the number of required alternations.

\end{abstract}

\begin{IEEEkeywords}
Nonlinear phase FIR design, minimax optimality, alternation theorem 
\end{IEEEkeywords}

\IEEEpeerreviewmaketitle

\section{Introduction}

Despite their desirable properties, FIR filters have certain disadvantages when compared to their IIR counterparts. For example, the minimum order required for an FIR filter to approximate a desired filter response within pre-specified bounds is usually much higher than the order of an IIR filter for the same task, which translates into more multiplications and additions per input sample in a hardware implementation and requires larger power and memory. Furthermore, even though linear phase FIR filters introduce no dispersion but only a uniform latency, the amount of this latency in samples is half the filter order and this may become unacceptably high for latency-sensitive applications. Therefore it is crucial to optimize an FIR filter for a given order while keeping this order as low as possible. The optimality can be stated with respect to a particular norm, and several design methods exist to meet different optimality criteria. For example, the windowing methods can be used to minimize the sum of absolute squares ($\l_2$-norm) of the approximation error between the filter response and the desired response for a given order. On the other hand, the Parks-McClellan method \cite{Parks1972} aims to minimize the maximum absolute error ($\l_{\infty}$-norm), also known as the minimax or Chebyshev error. Since minimizing the $l_2$-norm is equivalent to minimizing the energy in the error, it does not prevent large but very narrow deviations between the filter frequency response and the desired response that result in a small $\l_2$-norm. Therefore, even though it is much easier to design for $\l_2$-norm optimality, the minimax-optimality have been adopted widely in the signal processing community due to its superior worst-case performance, and it will be the focus in this paper.

The Parks-McClellan design method yields the unique globally-optimal linear phase FIR filter for a given order $N$ by exploiting a theorem that provides the necessary and sufficient conditions for global minimax-optimality, which is known as the alternation theorem \cite{Cheney1966}. Even though it is desirable, asserting linear phase restricts the coefficients of the filter to have even- or odd-symmetry if they are real-valued, or conjugate-symmetry if they are complex-valued. This restriction exhibits itself as halving the number of degrees of freedom available to approximate the desired response because each choice of a filter coefficient already determines another coefficient to generate symmetric pairs. For a variety of applications where linear phase is not crucial, this becomes an unnecessary constraint. Filters with magnitude responses that provide a much better approximation to the desired response can be obtained by removing the linear phase constraint. However, this renders the characterization of optimality by the alternation theorem inapplicable in its currently known form, which in turn prevents direct utilization of the Parks-McClellan method.

In this paper, we first provide a characterization method for the global minimax-optimality of FIR filters if no restrictions exist on its phase. In other words, we state the necessary and sufficient conditions for the magnitude response of an FIR filter to be the unique best approximation to a desired response. Of course, due to the restrictions that are imposed in a linear phase design, the optimal filter with unrestricted phase will always be at least as good an approximation to the desired response as the optimal linear phase solution. Since this is a characterization stated in terms of the magnitude response, all the FIR filters sharing the same order and magnitude response will be optimal. Therefore, although the magnitude response is the unique optimal response, there are a finite number of distinct optimal FIR filters related to each other through a cascade with an all-pass filter. Secondly, in this paper, the arguments of the characterization will be shown to naturally lead to a minimax-optimal design method involving the computation of an autocorrelation sequence as an intermediate step, at the end of which the designer will be able to choose from a variety of options for the phase including a minimum phase and a maximum phase design without compromising the global optimality of the magnitude response. Since it was originally introduced in Hermann and Schuessler's work \cite{Schuessler1970}, designing a nonlinear phase FIR filter by first designing an autocorrelation sequence and then finding a filter that admits this as its autocorrelation is a widely known technique \cite{Boite1981, Chen1986, Venkata1999, Kamp1983, Mian1982, Samueli1988, Wu1996}. Spectral factorization is an obvious first choice to obtain the filter coefficients from the designed autocorrelation sequence. This requires finding the roots of the polynomial the coefficients of which are the same as the autocorrelation sequence. Since this is a highly impractical approach for designing high order filters, for minimum phase designs, several algorithms have been proposed as an alternative to polynomial root finding \cite{Boite1981, Chen1986, Venkata1999, Kamp1983, Mian1982}. However, in most of these earlier methods for nonlinear phase FIR filter design, it is the autocorrelation sequence of the filter impulse response that is designed to have minimax optimality in approximating the desired response, which does not necessarily imply the optimality of the filter itself. Furthermore, since the magnitude response of the filter and that of its autocorrelation sequence are related through squaring, the weight function applied during the design of the autocorrelation will not match the desired weight function on the approximation error attained by the final design. In this paper, we first characterize the optimality of the nonlinear phase FIR filter instead of its autocorrelation sequence and then provide a method to compute the correct weight to be applied during the computation of the autocorrelation sequence so that the resulting filter exhibits the desired ratio of passband to stopband deviations. Spectral factorization using polynomial root finding or any of the alternative methods in the current literature \cite{Boite1981, Chen1986, Venkata1999, Kamp1983, Mian1982} can then be used to obtain the final design from the autocorrelation sequence leading to the FIR design with the desired error weights, where the global optimality is certified by our characterization of optimality.

\section{Linear Phase vs Nonlinear Phase FIR Filters}
Linear phase FIR filters with real-valued coefficients can be expressed as a real-valued amplitude function $A(\omega)$ multiplied by a linear phase term $e^{-j\alpha \omega}$. Since this linear phase corresponds to a time delay by $\alpha$ samples, a usual approach to designing linear phase FIR filters is to first design the zero-phase filter response $A(\omega)$, which is necessarily symmetric in time domain for it to be zero phase. Afterwards, the filter is time delayed until it is causal, which corresponds to multiplying with a phase of the form $e^{-j\alpha \omega}$.

Parks and McClellan \cite{Parks1972} exploited the fact that the frequency response $A(\omega)$ of a symmetric zero-phase filter $a[n]$ of even order $N$ can be expressed in terms of real sinusoids. More specifically, for example if $a[n]=a[-n]$, then
\begin{equation}\label{eqn:A(w)}
A(\omega) = \sum_{-N/2}^{N/2} a[n]e^{-j\omega n} = a[0] + \sum_{n=1}^{N/2}2a[n]cos(n\omega).
\end{equation}
In other words, $A(\omega)$ can be expressed as a linear combination of the basis functions $\{cos(n\omega),\; n=0,1,\dots,N/2\}$. In order to approximate an ideal filter response $D(\omega)$, we search for the optimal set of linear combination coefficients for these basis functions such that the maximum absolute error is minimized. The alternation theorem asserts that we need $N/2+2$ alternations for the optimal filter, and the Remez Exchange algorithm can be used to efficiently find this set of coefficients.

Since nonlinear phase filters cannot be time-shifted by any amount to exhibit a real-valued (zero-phase) response, and since the alternation theorem and the Remez Exchange Algorithm apply only to real-valued functions, they cannot be used to characterize or design minimax optimal nonlinear phase FIR filters directly.

\section{Characterization Theorem for Nonlinear Phase Filters}\label{sec:charac}
In this section, we describe a very straightforward method to characterize the global minimax optimality of a given nonlinear phase FIR filter $h[n],\;n=0,1,\dots,N$, where the optimality is implied for the magnitude response of this filter, $|H(e^{j\omega})|$, as compared to the desired filter response, $D(\omega)$, which is unity in the passband and zero in the stopband. In other words, by examining the magnitude response $|H(e^{j\omega})|$, we will be able to tell whether this is a filter the magnitude response of which is the best approximation to $D(\omega)$ in that no other magnitude response achievable with the same order can attain a smaller infinity norm on the weighted approximation error $W(\omega)(|H(e^{j\omega})|-D(\omega))$. We will be able to characterize optimality for both real-valued and complex-valued filter coefficients, and we do not require any symmetry in the coefficients, therefore it applies in the most general case. In the next section, the arguments of this characterization will enable us to find an efficient algorithm to design nonlinear phase FIR filters in cases where only the magnitude response is specified and the phase is not restricted. However we will still be able to choose among different available phase characteristics including, for example, a minimum phase design without compromising global optimality with respect to magnitude. 

Assume that an FIR filter with coefficients $h[n], \;n=0, 1, 2,\dots, N$ and a frequency response $H(e^{j\omega})$ is provided with the passband $\Omega_P$ and the stopband $\Omega_S$, both of which are closed subsets of $(-\pi,\pi]$, to approximate the desired magnitude response 
\begin{equation}\label{eqn:D}
D(\omega)=\left\{\begin{array}{lr}
1, & \omega \in \Omega_P\\
0,& \omega \in \Omega_S\\
\end{array} \right..
\end{equation}
Further assume that a desired weight function $W_{des}(\omega)$ is provided that expresses the relative emphasis on the error in the stopband as compared to the passband. More specifically,
\begin{equation}\label{eqn:W_des}
W_{des}(\omega)=\left\{\begin{array}{lr}
1, & \omega \in \Omega_P\\
K_{des},& \omega \in \Omega_S\\
\end{array} \right.,
\end{equation}
where $K_{des}$ is a scalar given as a part of the filter specifications. The weighted error function, and the bounds on passband and stopband errors are defined as 
\begin{equation}\label{eqn:weighted_error}
E_W(\omega) = W_{des}(\omega)\left(\left|H(e^{j\omega})\right|-D(\omega)\right),
\end{equation}
\begin{equation}\label{eqn:delta_P}
\delta_P = \max_{\omega\in(\Omega_P \cup \Omega_S)}\left|E_W(\omega)\right|
\end{equation}
and
\begin{equation}\label{eqn:delta_S}
\delta_S = \frac{\delta_P}{K_{des}},
\end{equation}
respectively.

\begin{thm}
\label{thm:main_res} $|H(e^{j\omega})|$ is the unique minimax-optimal magnitude response that can be attained by any FIR filter of order $N$ in order to approximate the ideal filter magnitude response $D(\omega)$ with a desired weight function $W_{des}(\omega)$ if and only if the \underline{adjusted} weighted error function
\begin{equation}\label{eqn:adjusted_weighted_error}
E'_W(\omega)=W'_{des}(\omega)\left(\left|H(e^{j\omega})\right|-D'(\omega)\right)
\end{equation}
exhibits at least $N+2$ alternations if the filter coefficients $h[n]$ are restricted to be real-valued, or at least $2N+2$ alternations if they are not restricted to be real-valued. Here, $W'_{des}(\omega)$ and $D'(\omega)$ are defined as
\begin{equation}\label{eqn:adjusted_W_des}
W'_{des}(\omega)=\left\{\begin{array}{lr}
1, & \omega \in \Omega_P\\
2K_{des},& \omega \in \Omega_S\\
\end{array} \right.,
\end{equation}
and
\begin{equation}\label{eqn:adjusted_D}
D'(\omega)=\left\{\begin{array}{lr}
1, & \omega \in \Omega_P\\
\frac{\delta_S}{2},& \omega \in \Omega_S\\
\end{array} \right..
\end{equation}
\end{thm}

Note that in the context of linear phase FIR filters where the alternation theorem is used for characterization, the optimality is characterized by counting the alternations in the weighted error function which is computed using the desired response $D(\omega)$ and the weight function $W_{des}(\omega)$. The characterization given here for nonlinear phase FIR filters is based on the adjusted weighted error function $E'_W$ computed using $D'(\omega)$ and $W'_{des}(\omega)$ as in equation (\ref{eqn:adjusted_weighted_error}). 

The formal proof of the characterization theorem (Theorem \ref{thm:main_res}) is given in \cite{DemirtasTSP2016} and will be excluded here for brevity. However, intuitively, the optimality of $h[n]$ can be related to the optimality of its autocorrelation sequence $p[n]$ as follows. Due to the specific choice of values in the stopband for $D'(\omega)$ and $W'_{des}(\omega)$, the number of alternations and the points at which alternations occur are the same for $|H(e^{j\omega})|$ and $P(e^{j\omega})=|H(e^{j\omega})|^2$. More specifically, if $|H(e^{j\omega})|$ attains its extreme value at a specific frequency and hence form an alternation in $E'_W(\omega)$, then the Fourier transform of the autocorrelation function $P(e^{j\omega})=|H(e^{j\omega})|^2$ will also attain its extremal value and form an alternation in a related weighted error function at the same frequency. Therefore, the number of required alternations in the magnitude response of the filter can be related to that of the autocorrelation sequence, which is zero-phase and in turn can be characterized for optimality using the traditional form of the alternation theorem.

The number of required alternations for filters with complex-valued coefficients are larger than that of filters with real-valued coefficients. This also is consistent with intuition because it reflects the additional degrees of freedom in choosing the filter coefficients by relaxing the constraint to be real-valued. More formally, the autocorrelation function of such filters also have complex-valued coefficients in general, and they exhibit conjugate-symmetry instead of the even-symmetry in the real case. This means that the autocorrelation sequence $p[n]$ of a filter $h[n]$ with complex-valued coefficients satisfy
\begin{equation}
p[n] = p^*[-n],\; n=0,1,2,\dots, N,
\end{equation}
or equivalently
\begin{equation}
p_{re}[n] = p_{re}[-n], \; n=0,1,2,\dots, N
\end{equation}
and 
\begin{equation}
p_{im}[n] = -p_{im}[-n], \; n=0,1,2,\dots, N,
\end{equation}
where the subscripts $re$ and $im$ refer to the real and imaginary parts, respectively. The frequency response of a conjugate-symmetric autocorreation sequence $p[n]$ can be represented as 
\begin{IEEEeqnarray}{lCl}\label{eqn:P(w)_complex}
P(e^{j\omega})&=&\sum_{n=-N}^N p[n]e^{-j\omega n}\nonumber\\
&=&p[0]+\sum_{n=1}^N p[n]e^{-j\omega n} + p^*[n]e^{j\omega n}\\
&=&p[0]+\sum_{n=1}^N 2p_{re}[n]cos(n\omega)+\sum_{n=1}^N 2p_{im}[n]sin(n\omega)\nonumber
\end{IEEEeqnarray}
This implies that the frequency response $P(e^{j\omega})$ can be represented as a linear combination of basis functions given as
\begin{equation}\label{eqn:cos_sin_basis}
\{cos(n\omega),\;n=0,1,2,\dots,N\}\cup\{sin(n\omega),\;n=1,2,\dots, N\}
\end{equation}
This means the flexibility in choosing the coefficients of $h[n]$ as complex-valued results in an addition of $N$ basis functions to the set of available functions to represent $P(e^{j\omega})$, potentially leading to smaller approximation errors as intuitively expected. Furthermore, the basis set in (\ref{eqn:cos_sin_basis}) also satisfies the Haar condition and therefore leads to a unique optimal solution \cite{Cheney1966}. These additional $N$ basis functions manifest themselves as an increase by $N$ in the number of required alternations to satisfy the traditional form of the alternation theorem for the design of the autocorrelation. Therefore, Theroem \ref{thm:main_res} applies to complex-valued filters with $2N+2$ alternations as opposed to $N+2$ alternations.

\begin{figure}
\centering
\subfloat[]{\includegraphics[scale =0.35] {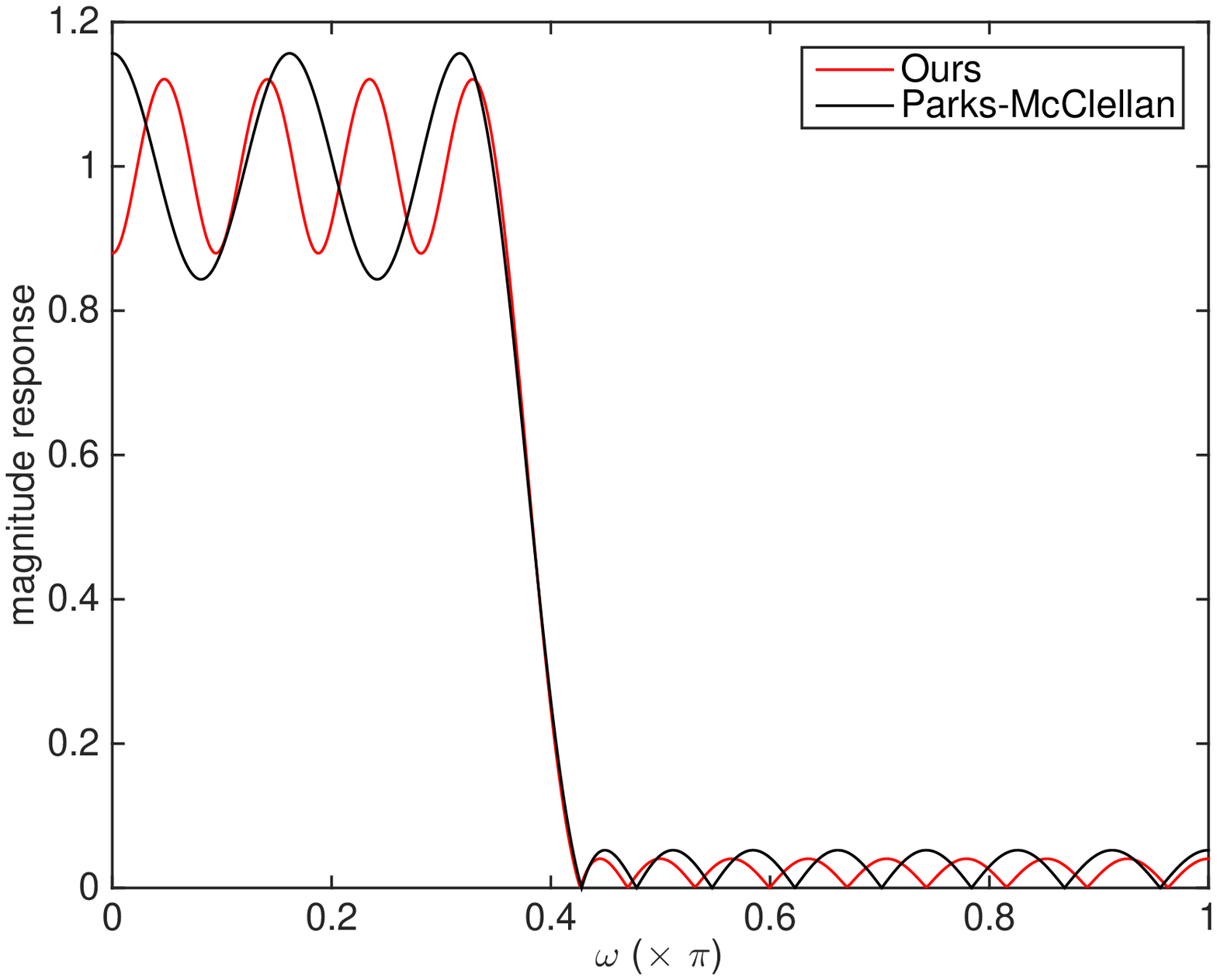}}\\
\subfloat[]{\includegraphics[scale =0.35] {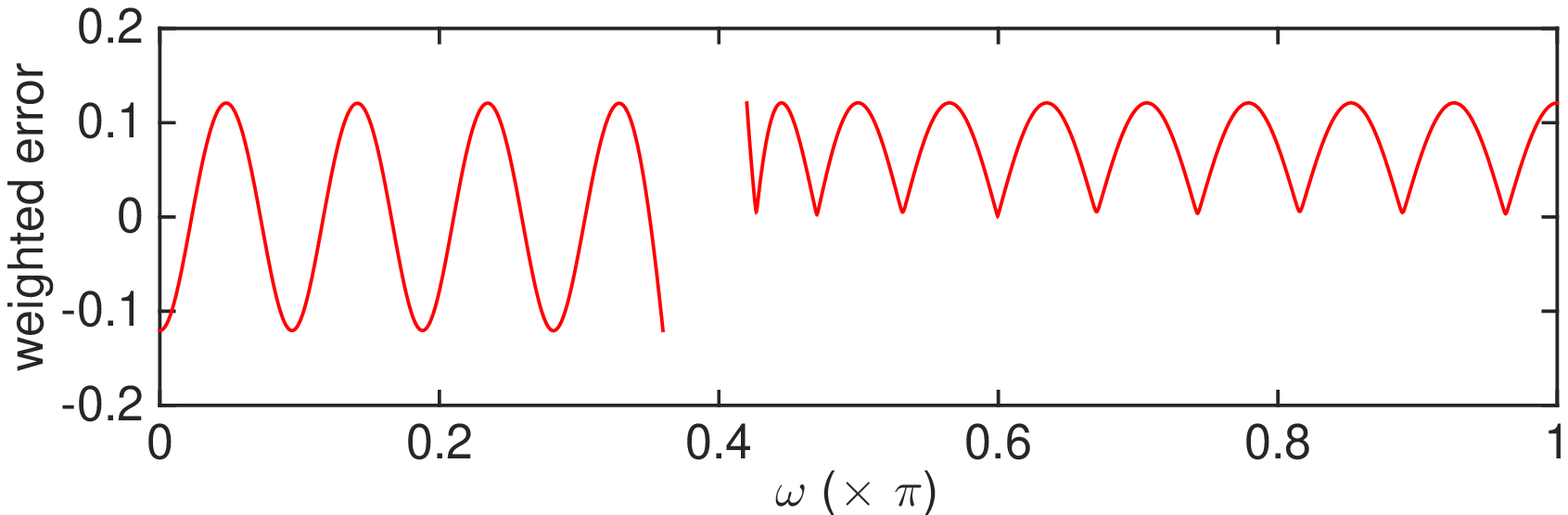}}\\
\subfloat[]{\includegraphics[scale =0.35] {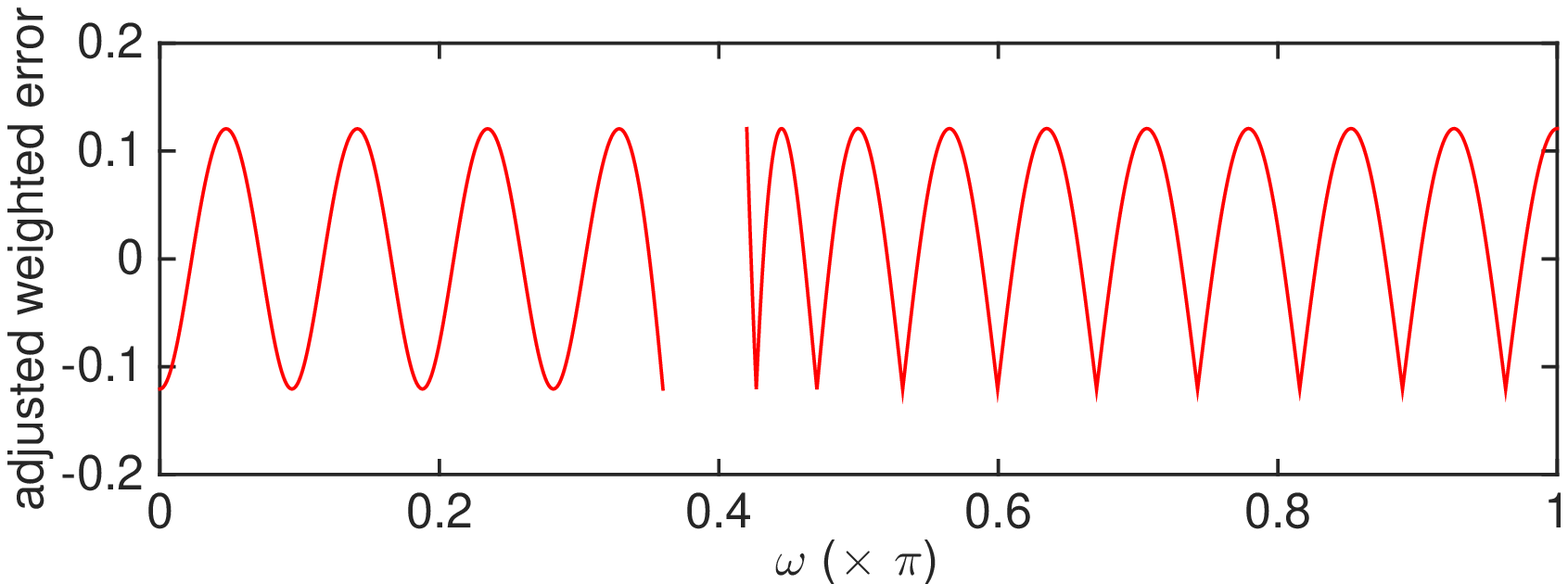}}\\
\subfloat[]{\includegraphics[scale =0.35] {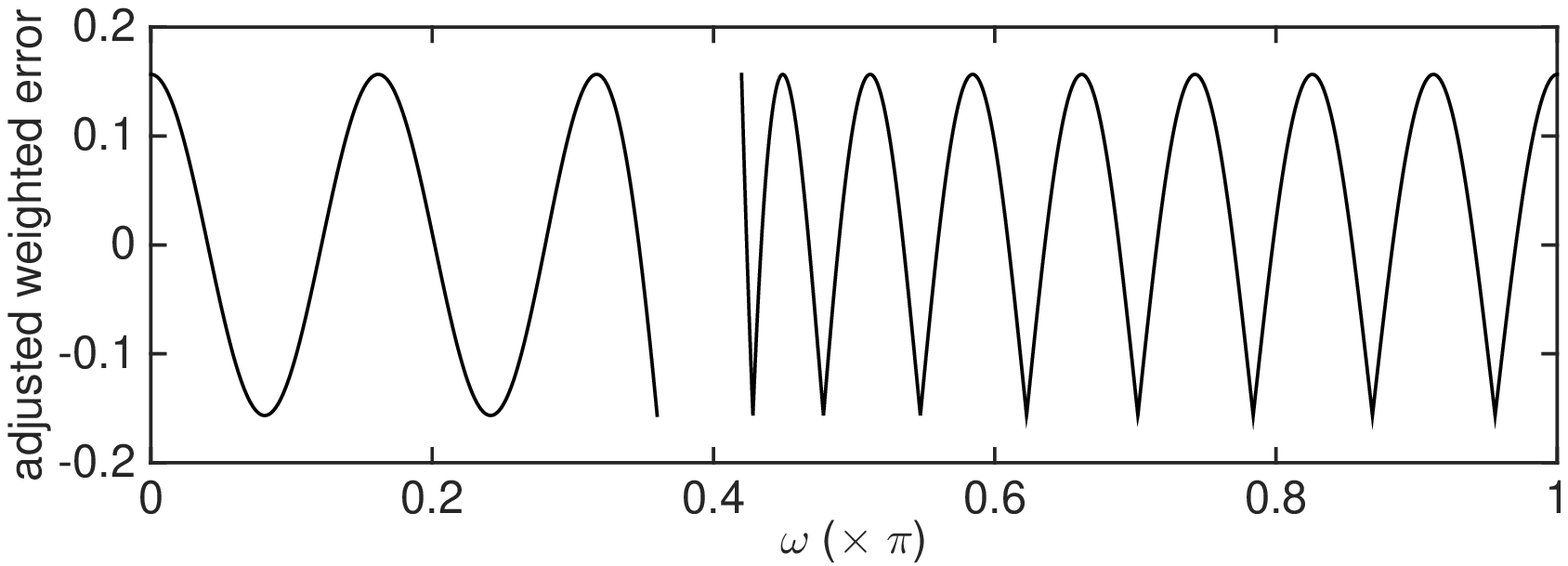}}
\caption{(a) The magnitude responses of two $26$-th order FIR filters with real-valued coefficients, one designed with the algorithm proposed in Section \ref{sec:design} and the other using \texttt{firpm} function of MATLAB (b) Weighted function $E_W(\omega)$ for our design (c) Adjusted weighted error $E'_W(\omega)$ for our design (d) Adjusted weighted error for the \texttt{firpm} design.}\label{fig:comparison}
\end{figure}

\textbf{Example:} Before proceeding to the design procedure, we close this section with an example of an FIR filter that is globally minimax optimal to illustrate the computation of the adjusted desired response $D'(\omega)$, the adjusted weight function $W'_{des}(\omega)$ and the alternation counting process in the characterization of nonlinear phase FIR filters for optimality. Our example design with real-valued coefficients in Figure \ref{fig:comparison}a has smaller ripple sizes than the one designed with MATLAB's \texttt{firpm} function which is based on the Parks-McClellan design \cite{Parks1972}. The filter order is $N=26$ with the passband and the stopband specified as $\Omega_P=[0,0.36\pi]$ and $\Omega_S=[0.42\pi,\pi]$. The weight $K_{des}$ is chosen as $3$ in this example, meaning the degrees of freedom will be chosen to suppress the stopband error more than the passband error by this factor. We compute $E_W(\omega)$ as in equation (\ref{eqn:weighted_error}), which is provided in Figure \ref{fig:comparison}b. From this computation, $\delta_P$ and $\delta_S$ are computed as $0.12$ and $0.04$, respectively. Therefore, $D'(\omega)$ and $W'_{des}(\omega)$ become
\begin{equation}
D'(\omega)=\left\{\begin{array}{lr}
1, & \omega \in [0,0.36\pi]\\
0.02,& \omega \in [0.42\pi,\pi]\\
\end{array} \right.
\end{equation}
and
\begin{equation}
W'_{des}(\omega)=\left\{\begin{array}{lr}
1, & \omega \in [0,0.36\pi]\\
6,& \omega \in [0.42\pi,\pi]\\
\end{array} \right..
\end{equation}
This leads to an adjusted weight function $E'_W(\omega)$ illustrated in Figure \ref{fig:comparison}c, which was obtained as in (\ref{eqn:adjusted_weighted_error}). This error indeed exhibits $N+2=28$ points, which is a necessary and sufficient condition for the unique and global optimality of the magnitude response $|H(e^{j\omega})|$ asserted by Theorem \ref{thm:main_res}. The \texttt{firpm} design exhibits only $23$ alternations in the adjusted weighted error which is computed similarly and illustrated in Figure \ref{fig:comparison}d; and is clearly suboptimal.

The formal steps provided for testing optimality can be bypassed by a more practical observation. Setting the maximum error in the passband to $\delta_P$, and the maximum error in the stopband to $\delta_S$, one can directly verify whether the filter satisfies the desired weight by checking if $\frac{\delta_P}{\delta_S}$ equals $K_{des}$. If this is the case, then the alternations can be also counted directly on the magnitude response as the points where the function reaches its extreme points in an alternating fashion including the band edges since $W'_{des}(\omega)$ and $D'(\omega)$ are tailored to turn these points into alternation points of the adjusted weighted error $E'_W(\omega)$. This approach can be verified in Figure \ref{fig:comparison}a where the extremal points are indeed alternations and lead to $28$ alternations including those occur at the band edges.

\section{Optimal Nonlinear Phase FIR Filter Design Algorithm}\label{sec:design}
In this section, we describe the design algorithm for filters restricted to having real-valued coefficients, therefore we will require $N+2$ alternations. The same arguments apply to filters with complex-valued coefficients simply by requiring $2N+2$ alternations and including sines in the basis functions for the computation of optimal squared response.
\subsection{Design Constraints}
Designing a zero-phase sequence $g[n]$ that approximates an ideal filter response, lifting its frequency response $G(e^{j\omega})$ until it is nonnegative and treating the lifted sequence $p[n]$ as the autocorrelation of an FIR filter $h[n]$ has been used as a nonlinear phase FIR filter design method at least since 1970 \cite{Schuessler1970}. However, since the design specifications such as relative weight on the stopband versus passband deviation in the autocorrelation domain do not remain the same for the filter due to the squaring relationship between $P(e^{j\omega})$ and $|H(e^{j\omega})|$, the resulting filter $h[n]$ does not necessarily reflect the desired weight. Furthermore, no optimality arguments are available for the final design $h[n]$ because the optimality of the autocorrelation sequence for one set of metrics does not make the corresponding filter optimal for the same metrics. We now provide a design method that correctly accounts for this relationship and computes the weight to be applied in the design of the autocorrelation sequence such that the final design exhibits the desired ratio between passband and stopband deviations. Furthermore, the characterization theorem of Section \ref{sec:charac} will certify the optimality of the filter itself as opposed to the optimality of the autocorrelation sequence.

Since the alternation frequencies and the number of alternations are the same for $|H(e^{j\omega})|$ and $P(e^{j\omega})$ due to the specific choice of $D'(\omega)$ and $W'_{des}(\omega)$ in the characterization theorem, designing the autocorrelation sequence $p[n]$ instead of the filter $h[n]$ itself with the correct number of alternations satisfies the conditions of the characterization theorem for the optimality of $h[n]$ itself. Therefore we can design an autocorrelation function that satisfies the required number of alternations and recover the filter coefficients that will accept this function as its autocorrelation function using either spectral factorization, or if a minimum phase filter is particularly desired, any of the methods in \cite{Boite1981, Chen1986, Venkata1999, Kamp1983, Mian1982}. 

\begin{figure}
\centering
\subfloat[]{\includegraphics[scale =0.35] {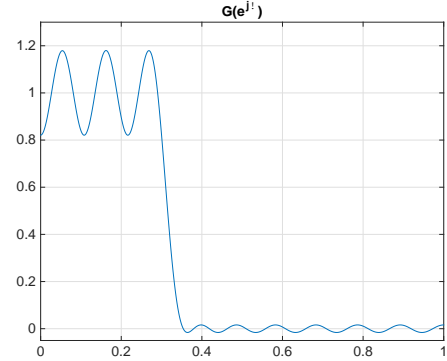}}\\
\subfloat[]{\includegraphics[scale =0.35] {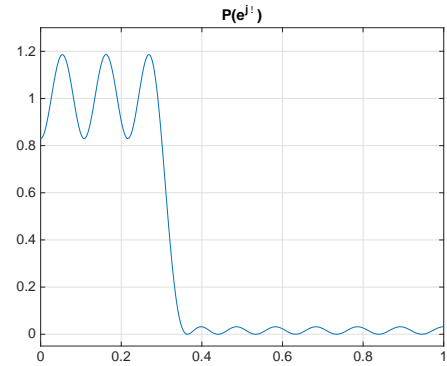}}\\
\subfloat[]{\includegraphics[scale =0.35] {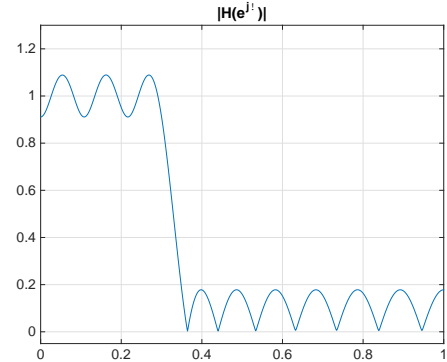}}\\
\caption{An example for (a) the Fourier Transform of the symmetric sequence $g[n]$ that approximates an ideal filter response, (b) that of the autocorrelation $p[n]$ obtained from $g[n]$ through scaling and shifting, i.e. $p[n]=ag[n]+b\delta[n]$, (c) the magnitude response of a nonlinear phase filter $h[n]$ the autocorrelation of which is $p[n]$. In this example, $N=20$, $\Omega_P=[0,0.30\pi]$, $\Omega_S=[0.35\pi,\pi]$ and $K_{des}=0.5$.}\label{fig:filter_and_autocorr}
\end{figure}

An autocorrelation sequence with at least $N+2$ alternations can be designed by first by computing the coefficients of an optimal even-symmetric sequence $g[n]$ of length $2N+1$ such that its Fourier transform $G(e^{j\omega})$ approximates the ideal filter, and then by scaling and shifting to obtain $P(e^{j\omega})=aG(e^{j\omega})+b$ such that the following constraints are satisfied:
\begin{itemize}
\item[(i)] $|H(e^{j\omega})|=\sqrt{P(e^{j\omega})}$ swings symmetrically around unity in the passband, i.e. its extremal values become $1+\delta_P$ and $1-\delta_P$ for some positive $\delta_P$,
\item[(ii)] the minimum value of $P(e^{j\omega})$ is zero,
\item[(iii)] the maximum value $\delta_S$ of $|H(e^{j\omega})|=\sqrt{P(e^{j\omega})}$ in the stopband satisfies the desired weight constraint, i.e., $\frac{\delta_P}{\delta_S}=K_{des}$.
\end{itemize}
The first condition guarantees that the effects of squaring are properly taken into account in the autocorrelation domain; the second constraint ensures $p[n]=ag[n]+b\delta[n]$ is a proper autocorrelation sequence; and the third constraint ensures that the desired weight is not compromised through squaring. The relationship between $G(e^{j\omega})$, $P(e^{j\omega})$ and $|H(e^{j\omega})|$ is illustrated with an example in Figure \ref{fig:filter_and_autocorr}. Referring to the passband and stopband deviations of $|H(e^{j\omega})|$ as $\delta_P$ and $\delta_S$ and those of $G(e^{j\omega})$ as $\Delta_P$ and $\Delta_S$ respectively, these three constraints can be represented mathematically in terms of the scale and shift coefficients $a$, $b$ and the weight $K$ that will be applied in the design of $g[n]$. More specifically, we choose the scaling coefficient $a$ and the shifting coefficient $b$ such that the midpoints of the passband and stopband ranges of $G(e^{j\omega})$ and $P(e^{j\omega})$ match as in
\begin{equation}
a\cdot 1 + b = 1 + \delta_P^2
\end{equation} 
and
\begin{equation}
a\cdot 0 + b = \frac{\delta_S^2}{2},
\end{equation} 
which yields 
\begin{equation}\label{eqn:a}
a = 1+\delta_P^2-\frac{\delta_S^2}{2}
\end{equation}
and
\begin{equation}\label{eqn:b}
b = \frac{\delta_S^2}{2}.
\end{equation}

The relative weight between passband and stopband does not change after scaling $G(e^{j\omega})$, therefore the weights are identical in both:
\begin{equation}
\frac{(1+\delta_P)^2-(1-\delta_P)^2}{\delta_S^2}=\frac{4\delta_P}{\delta_S^2}=\frac{\Delta_P}{\Delta_S}=K
\end{equation}
Since $\delta_P/\delta_S=K_{des}$, we can write this as
\begin{equation}
K = \frac{4K_{des}}{\delta_S}
\end{equation}
or
\begin{equation}\label{eqn:delta_S}
\delta_S = \frac{4K_{des}}{K}.
\end{equation}
In order to match the upper bound of the filter response in the stopband to that of the autocorrelation after the scale and shift, we have
\begin{equation}\label{eqn:delta_S2}
\delta_S^2 = a\cdot \Delta_S + b.
\end{equation}
Inserting the values of $a$ and $b$ from equations (\ref{eqn:a}) and (\ref{eqn:b}), and inserting $\delta_S^2=\frac{16K_{des}^2}{K^2}$ from equation (\ref{eqn:delta_S}), we obtain
\begin{equation}
\frac{16K_{des}^2}{K^2}=\left(1+\frac{16K_{des}^4}{K^2}-\frac{8K_{des}^2}{K^2}\right)\Delta_S+\frac{8K_{des}^2}{K^2}
\end{equation}
Solving this for $\Delta_S$ yields
\begin{equation}\label{eqn:Delta_S}
\Delta_S = \frac{8K_{des}^2}{K^2+16K_{des}^4-8K_{des}^2}.
\end{equation}
and, since $\Delta_P = K\Delta_S$, we obtain
\begin{equation}\label{eqn:Delta_P}
\Delta_P = \frac{8K_{des}^2K}{K^2+16K_{des}^4-8K_{des}^2}.
\end{equation}

Finally, once the appropriate weight $K$ that satisfies this equation is found, the scaling and shifting coefficients can be computed directly from the parameters of this filter. Using equation (\ref{eqn:delta_S}) and (\ref{eqn:b}), we obtain
\begin{equation}\label{eqn:b_better}
b = \frac{8K_{des}^2}{K^2},
\end{equation}
and from equation (\ref{eqn:b_better}) and (\ref{eqn:delta_S2}), we have
\begin{equation}\label{eqn:a_better}
a = \frac{8K_{des}^2}{K^2\Delta_S} = \frac{8K_{des}^2}{K\Delta_P}.
\end{equation}

Equation (\ref{eqn:Delta_P}) is an implicit and nonlinear equation that expresses the correct weight $K$ that needs to be applied in the design of $G(e^{j\omega})$ in terms of the actual passband deviation $\Delta_P$ obtained using the Remez Exchange Algorithm. It can be solved efficiently for example using an iterative procedure where we cut the search space on $K$ each time in a binary search fashion or using Newton-Raphson method. Once the appropriate $K$ is found, we design $G(e^{j\omega})$ with that weight, scale by $a$ and shift by $b$ to obtain the autocorrelation response $P(e^{j\omega})$, and then recover the filter coefficients $h[n]$ using spectral factorization where we can choose the filter to be minimum phase, or maximum phase or anything in between. If the desired filter is minimum phase, then more efficient methods in \cite{Boite1981, Chen1986, Venkata1999, Kamp1983, Mian1982} can be used instead of spectral factorization.

\subsection{The Overall Algorithm}\label{subsec:overall}
Assume that we want to compute the minimax-optimal magnitude response $|H(e^{j\omega})|$, which is of order $N$ (therefore has $N+1$ coefficients). Given $\Omega_S$, $\Omega_P$ and $K_{des}$, start with an initial guess for $K$ such that $K\ge4K_{des}(K_{des}+1)$ for a physically meaningful design. (See Appendix \ref{app:unique_intersection} for the derivation of this lower bound for weight as well as the justification for why the following iterations converge.)
\begin{itemize}
\item[1.] Compute the coefficients of the minimax-optimal even-symmetric (therefore zero-phase) filter $g[n]$ of order $2N$ to approximate the target function
\begin{equation}
D(\omega)=\left\{\begin{array}{lr}
1, & \omega \in \Omega_P\\
0,& \omega \in \Omega_S\\
\end{array} \right.
\end{equation}
with the weight function
\begin{equation}
W(\omega)=\left\{\begin{array}{lr}
1, & \omega \in \Omega_P\\
K,& \omega \in \Omega_S\\
\end{array} \right..
\end{equation}
This can be done directly using the Remez Exchange algorithm, or modifying the Parks-McClellan algorithm, etc.
\item[2.] Compute $G(e^{j\omega})$, the frequency response of $g[n]$, and compute the maximum value of passband error $\Delta_P$ for this $g[n]$. (This is also equivalent to the maximum value of the absolute weighted error $|W(\omega)(G(e^{j\omega})-D(\omega))|$).
\item[3.] If the resulting $\Delta_P$ from step 2 satisfies the equality in equation (\ref{eqn:Delta_P}), go to step 4. Otherwise
\begin{itemize}
\item[(i)] If $\Delta_P$ is smaller than the expression in equation (\ref{eqn:Delta_P}), then increase the value of $K$ and go to step 1.
\item[(i)] If $\Delta_P$ is greater than the expression in equation (\ref{eqn:Delta_P}), then decrease the value of $K$ and go to step 1.
\end{itemize}
The amount by which $K$ is increased or decreased, or the bounds of the search space, can be decided in several ways, including methods such as binary search (or bisection method), Newton-Raphson method, or any other appropriate numerical method. Any choice of $K$ must satisfy $K\ge4K_{des}(K_{des}+1)$ for a physically meaningful design. 
\item[4.] Compute scale and shift coefficients $a$ and $b$ using equations (\ref{eqn:b_better}) and (\ref{eqn:a_better}).
\item[5.] Compute the function $p[n]$ from $g[n]$ as $p[n]=a\cdot g[n]+b\cdot \delta[n]$, where $\delta[n]$ is the unit impulse function (not to be confused with passband or stopband ripples of $|H(e^{j\omega})|$, namely $\delta_P$ or $\delta_S$.). This $p[n]$ is the autocorrelation that we were looking for.
\item[6.] Using any method including but not limited to spectral factorization, obtain the coefficients of $h[n]$ from the autocorrelation sequence $p[n]$. There will be more than one filter for which the autocorrelation sequence is $p[n]$, all of which are related to one another through a cascade with an allpass filter. Choosing the zeros in the unit circle as well as one from each pair of zeros located on the unit circle leads to a minimum phase design. Choosing all the zeros outside the unit circle and one pair from each pair on the unit circle leads to a maximum phase design. If a minimum phase design is desired but the filter order is too high to perform spectral factorization using polynomial root finding, methods in \cite{Boite1981, Chen1986, Venkata1999, Kamp1983, Mian1982} can be used to find the minimum phase solution. Other solutions can then be found by finding the roots of this lower order polynomial and reflecting the zeros from inside the unit circle to outside of the unit circle as desired.
\end{itemize}

\begin{figure}
\centering
\subfloat[]{\includegraphics[scale =0.25] {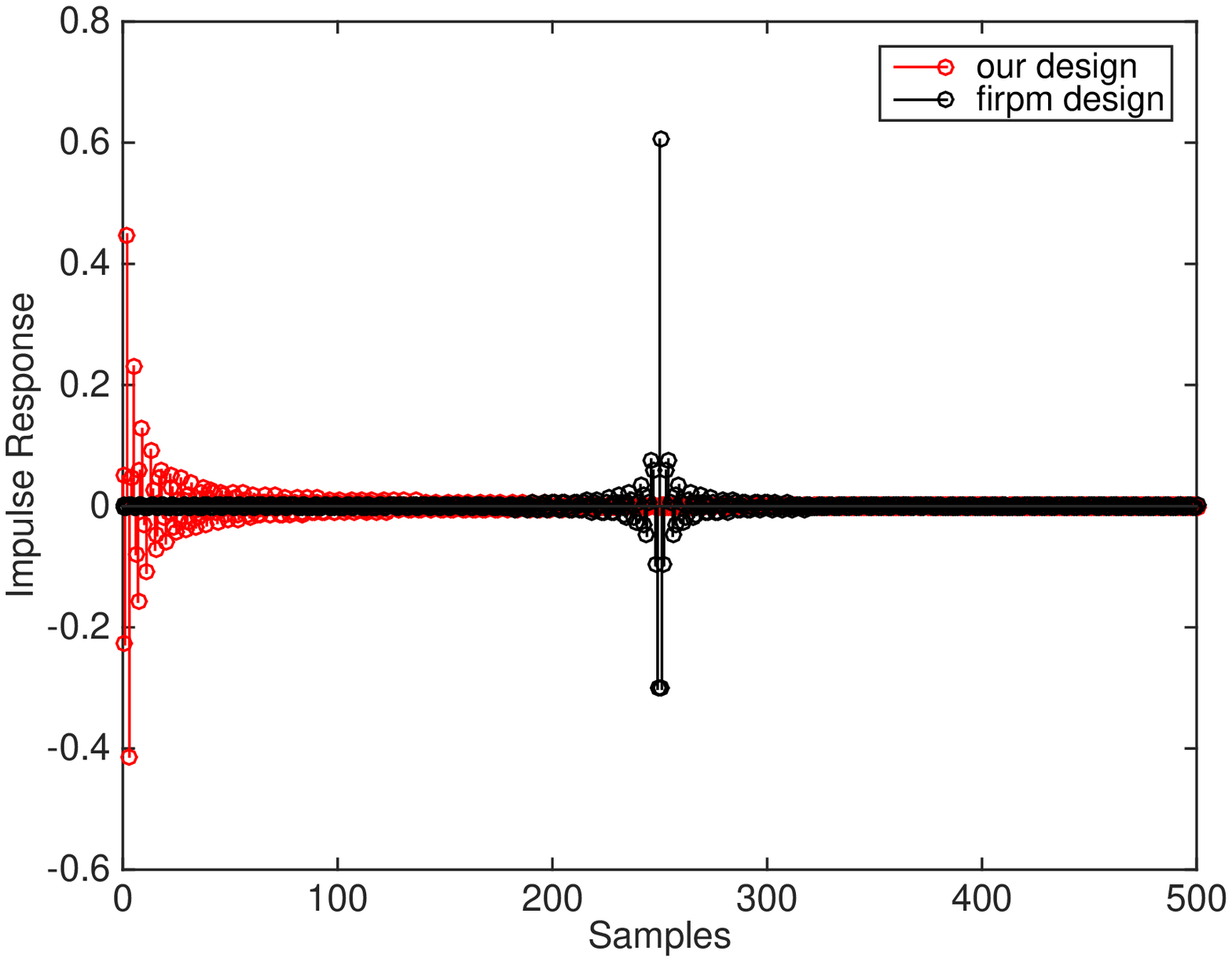}}\\
\subfloat[]{\includegraphics[scale =0.25] {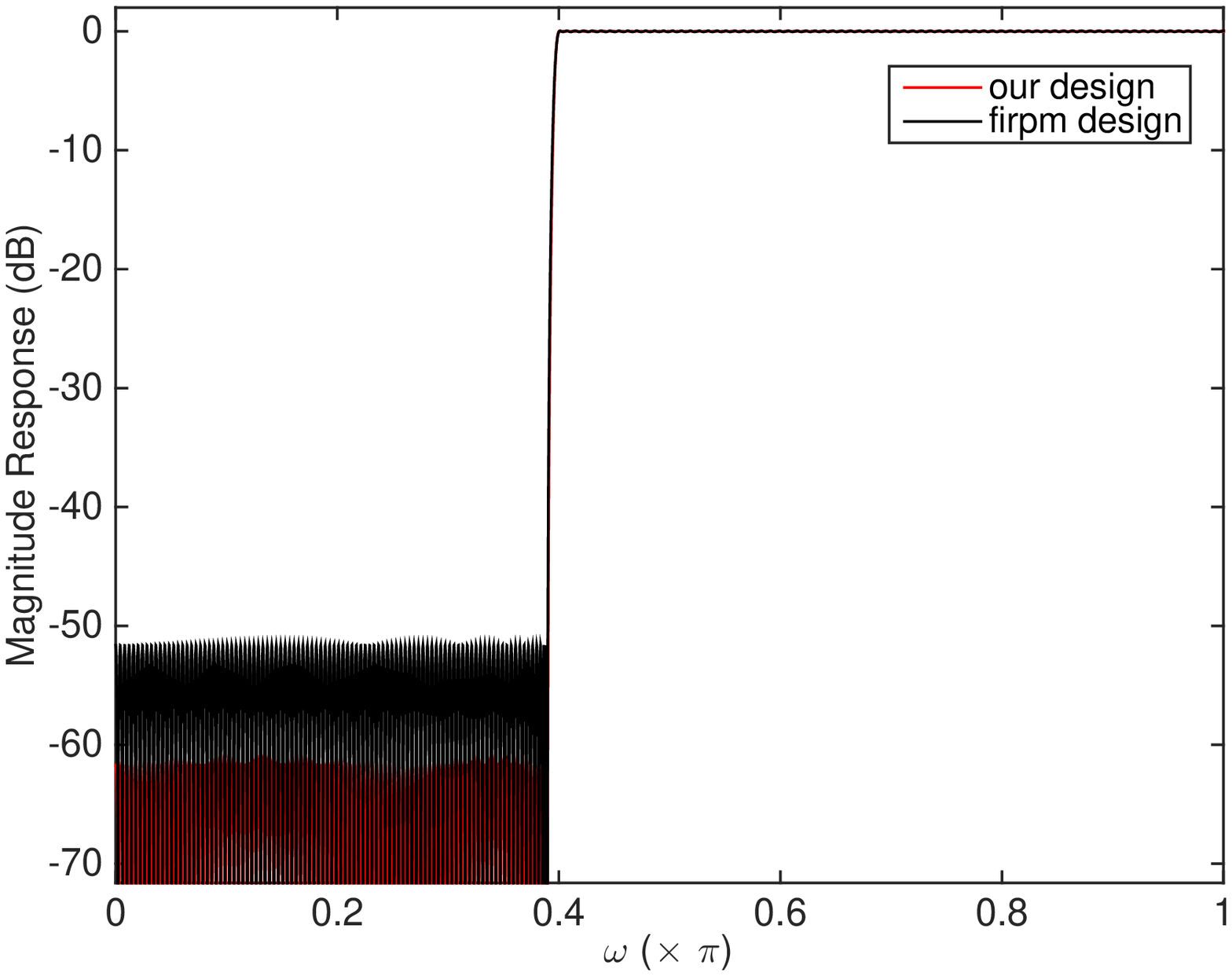}}\\
\subfloat[]{\includegraphics[scale =0.25] {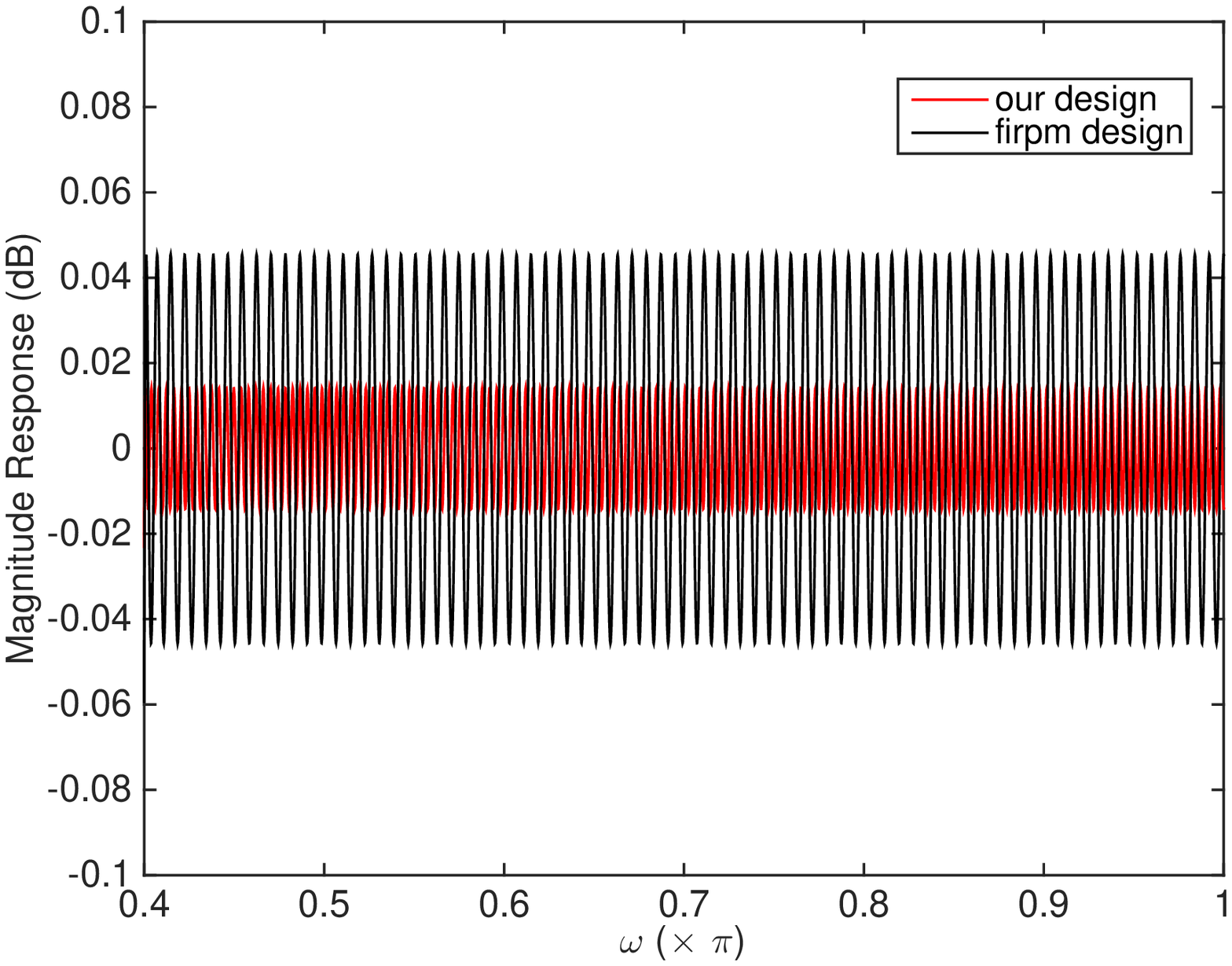}}\\
\subfloat[]{\includegraphics[scale =0.25] {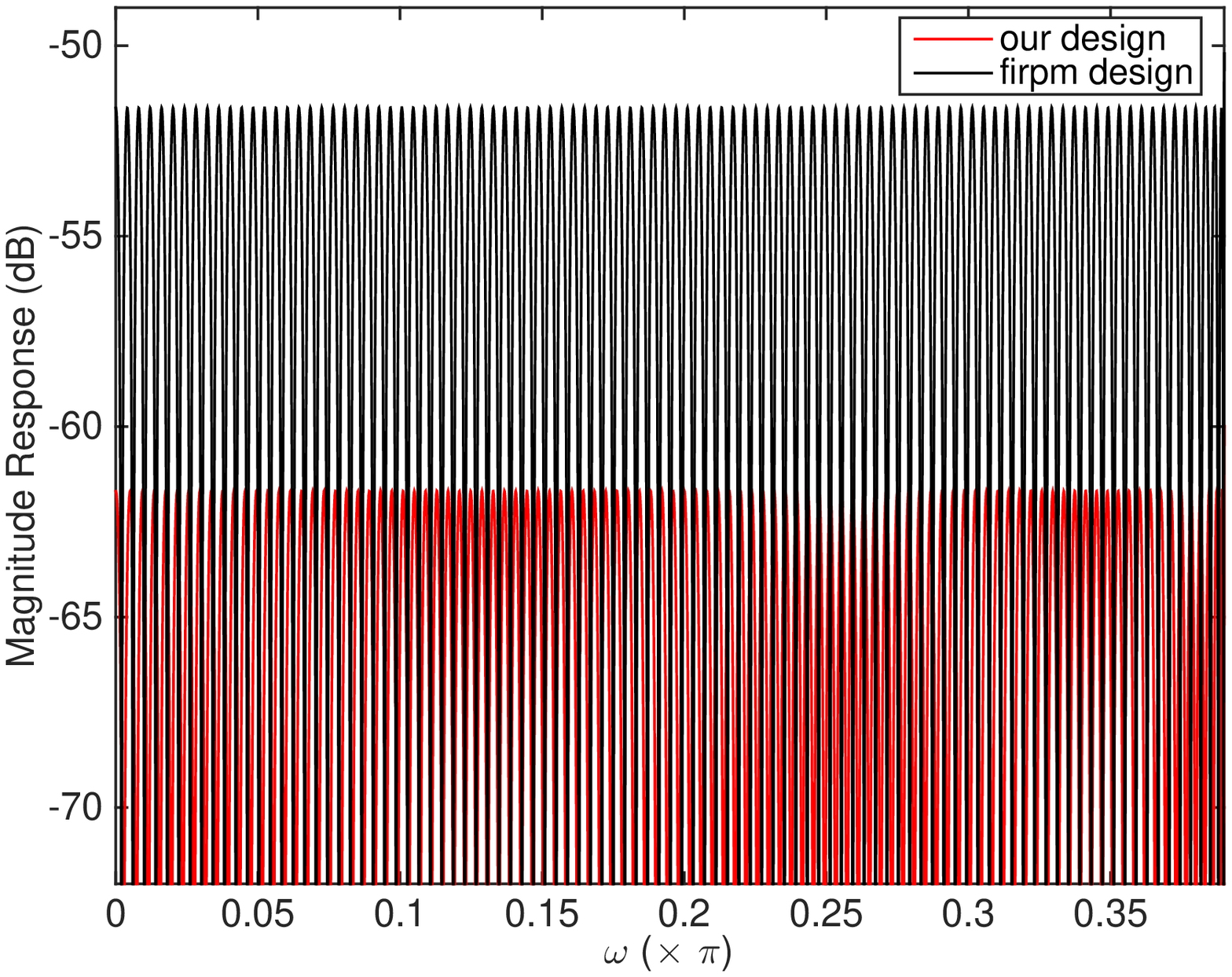}}\\
\subfloat[]{\includegraphics[scale =0.25] {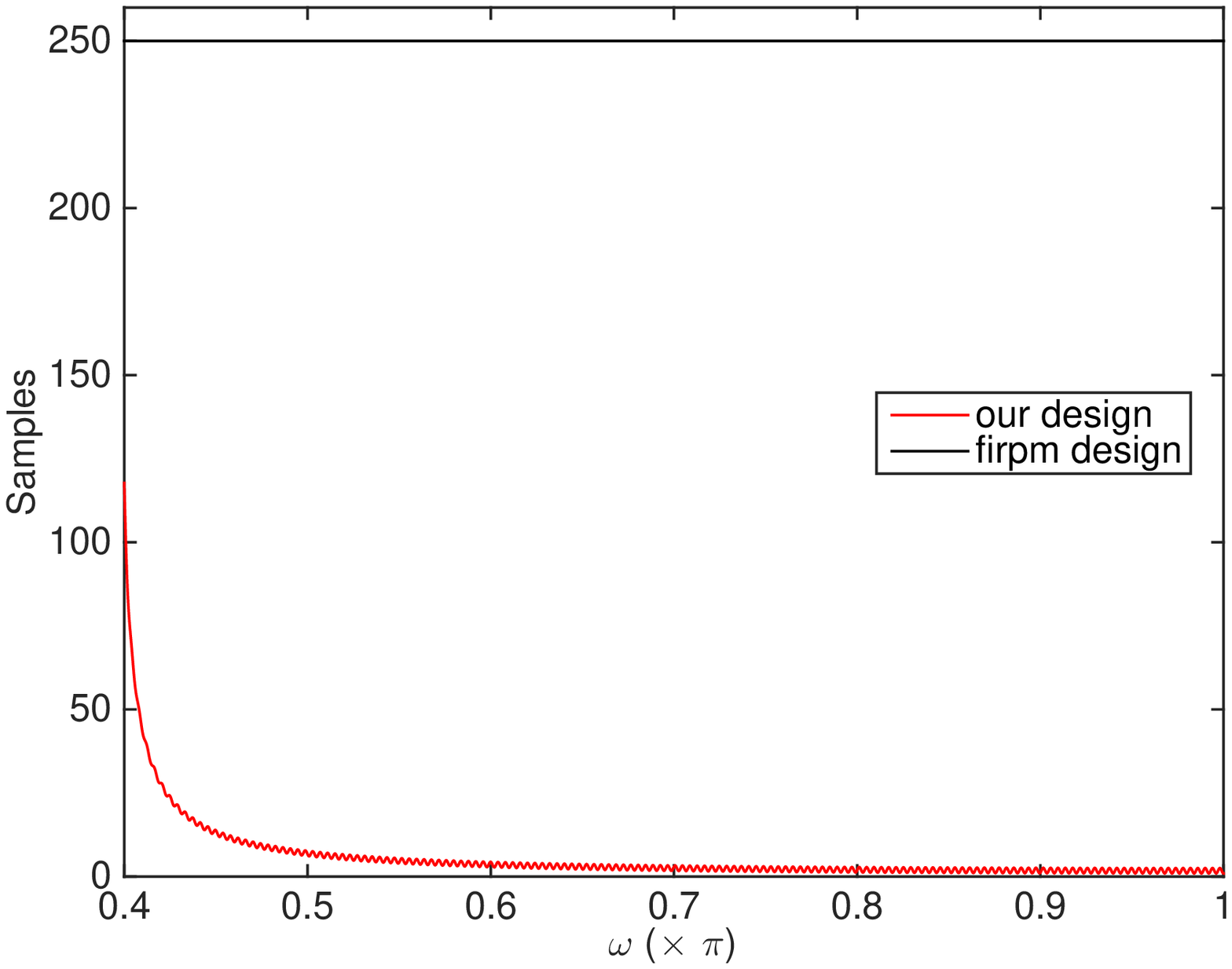}}\\
\caption{Comparison of our optimal design to a linear phase design for a high-pass filter where $N=500$, $\Omega_P=[0.40\pi, \pi]$, $\Omega_S=[0,0.39\pi]$ and $K_{des}=2$. (a) the impulse responses, (b) the magnitude responses, (c) the magnitude response zooming into the passband, (d) the magnitude response zooming into the stopband, (e) the group delays in samples in the passband. }\label{fig:high_order_example}
\end{figure}

\textbf{Example:} Figure \ref{fig:high_order_example} illustrates the design of a high pass filter where $N=500$, $\Omega_P=[0.40\pi, \pi]$, $\Omega_S=[0,0.39\pi]$ and $K_{des}=2$. In this particular design, for the desired relative ratio of $K_{des}=2$ between the passband and the stopband deviations, the weight $K$ to be applied during the the design of the zero phase sequence $g[n]$ was computed as $9801.96$ corresponding to $\Delta_P=3.2646E-3$ in the iterative procedure in Section \ref{subsec:overall}. This pair of $(K, \Delta_P)$ can be verified to satisfy equation (\ref{eqn:Delta_P}) which was obtained through only few iterations using the bisection method.

Figure \ref{fig:high_order_example}a illustrates the minimum phase impulse response obtained using our design algorithm, which is the globally optimal solution that approximates the desired response if no restrictions on phase exist as certified by Theorem \ref{thm:main_res}, and that is designed as a linear phase filter using the \texttt{firpm} function in MATLAB based on the Parks-McClellan algorithm \cite{Parks1972}. Figure \ref{fig:high_order_example}b, \ref{fig:high_order_example}c and \ref{fig:high_order_example}d illustrate the comparisons of magnitude responses in the entire frequency range, in the passband and in the stopband respectively. Removing the restrictions on the phase has clearly led to a sharper magnitude response characteristic. Furthermore, since this is a minimum phase design, the group delay in the entire passband is much less than that of the linear phase design at the expense of a non-constant delay profile. Since this is a very high order filter, a polynomial root finding based spectral factorization was impractical and we exploited the algorithm in 
\cite{Boite1981} instead to compute $h[n]$ from the computed $p[n]$, where we used a matlab implementation of this algorithm provided in \cite{JOSweb}.

\appendices

\section{Proof for Convergence of Iterations}\label{app:unique_intersection}
First we show that the resulting passband ripple $\Delta_{P,res}$ of a zero-phase $G(e^{j\omega})$ obtained by applying a weight $K$ is an increasing function of $K$, i.e. if we design a filter for the same passband and stopband specifications but with a larger $K$, then $\Delta_{P,res}$ will be also higher. Here the subscript $res$ stands for ``resulting". We can prove that $\Delta_{P,res}(K)$ is an increasing function of $K$ by contradiction. 

Assume that the maximum weighted error in a filter $G_1(e^{j\omega})$ designed with a weight function
\begin{equation}
W_1(\omega)=\left\{\begin{array}{lr}
1, & \omega \in \Omega_P\\
K_1,& \omega \in \Omega_S\\
\end{array} \right.
\end{equation}
is given by
\begin{equation}
\Delta_{P1} = \max_{\omega\in(\Omega_P \cup \Omega_S)}\left|E_{W1}(\omega)\right|
\end{equation}
where
\begin{equation}
E_{W1}(\omega) = W_1(\omega)\left(G_1(e^{j\omega})-D(\omega)\right).
\end{equation}
Similarly define $W_2(\omega)$, $E_{W2}(\omega)$ and $\Delta_{P2}$ for another filter $G_2(e^{j\omega})$ and weight scalar $K_2$, which is greater than $K_1$. Since $G_1$ and $G_2$ are real-valued frequency responses, they can be designed using the Remez Exchange Algorithm and they will be the respective unique optimal-minimax designs since they satisfy the alternation theorem with at least $N+2$ alternations.

In order to obtain the contradiction, assume $\Delta_{P2}\le \Delta_{P1}$. We can compute another weighted error function $E_{12}(\omega)$ using the weight $W_1(\omega)$ along with $G_2(e^{j\omega})$ as
\begin{equation}
E_{12}(\omega) = W_1(\omega)\left(G_2(e^{j\omega})-D(\omega)\right).
\end{equation} 
The maximum value of this weighted error $\Delta_{P12}$ becomes
\begin{equation}
\Delta_{P12}=\max_{\omega\in(\Omega_P \cup \Omega_S)}\left|E_{12}(\omega)\right| = \max\left\{\Delta_{P2},K_1\Delta_{S2} \right\}.
\end{equation}
Since both $\Delta_{P2}$ and $K_1\Delta_{S2}=K_1\frac{\Delta_{P2}}{K_2}$ are at most $\Delta_{P2}$, the filter $G_2(e^{j\omega})$ is at least as good as $G_1(e^{j\omega})$ in approximating the desired function with the weight function $W_1(\omega)$, which contradicts the unique optimality of the latter which had been established by the alternation theorem. Therefore, for $K_2>K_1$, the passband error $\Delta_{P2}$ cannot be equal or smaller than $\Delta_{P1}$. This implies $\Delta_P(K)$ is a strictly increasing function of $K$.

Now consider the range of weights $K$ that results in physically meaningful designs. For a given weight $K_{des}$, we can always design a trivial filter $h[n]=\frac{1}{1+K_{des}}\delta[n]$, which leads to stopband error $\delta_S=\frac{1}{1+K_{des}}$ and passband error $\delta_P=\frac{K_{des}}{1+K_{des}}$, which satisfy the desired weight constraint $\frac{\delta_P}{\delta_S}=K_{des}$. The sum of these errors is exactly unity. Therefore, in an optimal design with $N+1$ coefficients, the sum of these errors can never be larger than unity, and we have
\begin{equation}
\delta_P+\delta_S = (K_{des}+1)\delta_S \le 1.
\end{equation}
Inserting $\delta_S$ from equation (\ref{eqn:delta_S}), we obtain
\begin{equation}
(K_{des}+1)\frac{4K_{des}}{K}\le 1,
\end{equation}
which yields the physically meaningful lower bound for the weight $K$ as
\begin{equation}\label{eqn:physically_meaningful_K}
K\ge4K_{des}(K_{des}+1).
\end{equation}

\begin{figure}
\begin{center}
\includegraphics[scale =0.35] {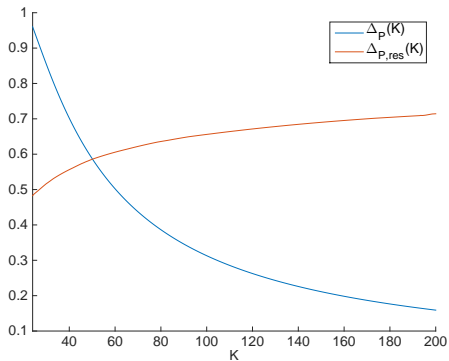}
\caption{An example of the resulting passband deviation $\Delta_{P,res}$ as a function of weight K and the function $\Delta_P$ given in equation (\ref{eqn:Delta_P}).}\label{fig:unique_intersect}
\end{center}
\end{figure}

We need to find the weight $K$ that yields a resulting value of weighted error $\Delta_{P,res}$ that is equal to the expression for $\Delta_P$ in equation (\ref{eqn:Delta_P}). We have already established $\Delta_{P,res}(K)$ is an increasing function of $K$ and is always lower than $1-\Delta_S(K)$. On the other hand, $\Delta_P(K)$ stated in equation (\ref{eqn:Delta_P}) equals $1-\delta_S$ at the lower bound we just found in equation (\ref{eqn:physically_meaningful_K}), is monotonically decreasing for weights greater than this lower bound and asymptotically approaching zero. This means these two functions do intersect once and only once in this regime of weights, and that is the weight we are looking for. This is also another proof that an optimum exists and it is unique. Figure \ref{fig:unique_intersect} illustrates an example of $\Delta_P(K)$ and $\Delta_{P,res}$ curve for $w_{des}=2$, $\Omega_S=[0,0.36\pi]$ and $\Omega_P=[0.42\pi,\pi]$ where the physically meaningful values of $K$ start as $24$. During the iterations, if $\Delta_{P,res}(K)>\Delta_P(K)$, the current value of $K$ is to the right of the intersection and therefore needs to be decreased. Otherwise, it is increased, and the iterations continue until the two values meet. The search is particularly efficient if the search space is halved each time, corresponding to the binary search (or bisection) scheme.

\ifCLASSOPTIONcaptionsoff
  \newpage
\fi





\bibliographystyle{IEEEtran}
\bibliography{IEEEabrv,/Users/sdemirta/Documents/Dropbox/Personal/new_research/my_references}{}

\begin{thebibliography}{10}
\providecommand{\url}[1]{#1}
\csname url@samestyle\endcsname
\providecommand{\newblock}{\relax}
\providecommand{\bibinfo}[2]{#2}
\providecommand{\BIBentrySTDinterwordspacing}{\spaceskip=0pt\relax}
\providecommand{\BIBentryALTinterwordstretchfactor}{4}
\providecommand{\BIBentryALTinterwordspacing}{\spaceskip=\fontdimen2\font plus
\BIBentryALTinterwordstretchfactor\fontdimen3\font minus
  \fontdimen4\font\relax}
\providecommand{\BIBforeignlanguage}[2]{{%
\expandafter\ifx\csname l@#1\endcsname\relax
\typeout{** WARNING: IEEEtran.bst: No hyphenation pattern has been}%
\typeout{** loaded for the language `#1'. Using the pattern for}%
\typeout{** the default language instead.}%
\else
\language=\csname l@#1\endcsname
\fi
#2}}
\providecommand{\BIBdecl}{\relax}
\BIBdecl

\bibitem{Parks1972}
T.~Parks and J.~McClellan, ``Chebyshev approximation for nonrecursive digital
  filters with linear phase,'' \emph{Circuit Theory, IEEE Transactions on},
  vol.~19, no.~2, pp. 189 -- 194, Mar 1972.

\bibitem{Cheney1966}
E.~W. Cheney, \emph{Introduction to approximation theory}.\hskip 1em plus 0.5em
  minus 0.4em\relax McGraw-Hill, 1966.

\bibitem{Schuessler1970}
O.~Herrmann and W.~Schuessler, ``Design of nonrecursive digital filters with
  minimum phase,'' \emph{Electronics Letters}, vol.~6, no.~11, pp. 329--330,
  May 1970.

\bibitem{Boite1981}
R.~Boite and H.~Leich, ``A new procedure for the design of high order minimum
  phase fir digital or ccd filters.'' \emph{Signal Processing}, vol.~3, pp. 101
  -- 108, 1981.

\bibitem{Chen1986}
X.~Chen and T.~W. Parks, ``Design of optimal minimum phase fir filters by
  direct factorization,'' \emph{Signal Process.}, vol.~10, no.~4, pp. 369--383,
  Jun. 1986.

\bibitem{Venkata1999}
N.~Damera-Venkata and B.~L. Evans, ``Optimal design of real and complex minimum
  phase digital fir filters,'' in \emph{Acoustics, Speech, and Signal
  Processing, 1999. Proceedings., 1999 IEEE International Conference on},
  vol.~3, Mar 1999, pp. 1145--1148 vol.3.

\bibitem{Kamp1983}
Y.~Kamp and C.~Wellekens, ``Optimal design of minimum-phase fir filters,''
  \emph{IEEE Transactions on Acoustics, Speech, and Signal Processing},
  vol.~31, no.~4, pp. 922--926, Aug 1983.

\bibitem{Mian1982}
G.~Mian and A.~Nainer, ``A fast procedure to design equiripple minimum-phase
  fir filters,'' \emph{IEEE Transactions on Circuits and Systems}, vol.~29,
  no.~5, pp. 327--331, May 1982.

\bibitem{Samueli1988}
H.~Samueli, ``On the design of optimal equiripple fir digital filters for data
  transmission applications,'' \emph{IEEE Transactions on Circuits and
  Systems}, vol.~35, no.~12, pp. 1542--1546, Dec 1988.

\bibitem{Wu1996}
S.-P. Wu, S.~Boyd, and L.~Vandenberghe, ``Fir filter design via semidefinite
  programming and spectral factorization,'' in \emph{Decision and Control,
  1996., Proceedings of the 35th IEEE Conference on}, vol.~1, Dec 1996, pp.
  271--276 vol.1.

\bibitem{DemirtasTSP2016}
S.~Demirtas, ``Characterization and design of nonlinear phase fir filters for
  global minimax-optimality,'' \emph{in preparation}.

\bibitem{JOSweb}
J.~O. Smith, \emph{Introduction to Digital Filters with Audio
  Applications}.\hskip 1em plus 0.5em minus 0.4em\relax
  \htmladdnormallink{\texttt{http:}}{http://ccrma.stanford.edu/~jos/filters/}\texttt{//\-ccrma.stanford.edu/\-\~{}jos/\-filters/},
  accessed (May 2016), online book.

\end{thebibliography}
\end{document}